\documentclass[cameraready]{Interspeech}

\title{Beyond Mimicry: Constrained Exploration with GRPO for Joint Multi-Talker ASR and Diarization under Unknown Speaker Counts}

\author[affiliation={1}]{Yunrui}{Cai}
\author[affiliation={1}]{Dingdong}{Wang}
\author[affiliation={1}]{Lingwei}{Meng}
\author[affiliation={1}]{Xixin}{Wu}
\author[affiliation={1,2}]{Zhiyong}{Wu}
\author[affiliation={1},correspondingauthor]{Helen}{Meng}

\address{
    $^1$ The Chinese University of Hong Kong, Hong Kong SAR, China \\
    $^2$ Shenzhen International Graduate School, Tsinghua University, Shenzhen, China
}

\email{\{yrcai,ddwang,lmeng,wuxx,zywu,hmmeng\}@se.cuhk.edu.hk}


\usepackage{comment}
\usepackage{multirow}

\begin{document}

\maketitle

\begin{abstract}
Joint multi-talker ASR with speaker attribution and timestamps is particularly challenging when the speaker count is unknown and the output must follow a strict structure. We observe that Supervised Fine-Tuning (SFT)-only alignment of SpeechLLMs is brittle under dense overlaps, often leading to burst hallucinations and malformed speaker/timestamp tags. We propose a unified two-stage generative framework: (i) CoT-augmented SFT that infers the global speaker count before transcription, and (ii) GRPO-based constrained exploration guided by a Multi-dimensional Constraint-Aware Reward (MCAR) that directly optimizes permutation-invariant accuracy while enforcing counting, temporal, and structural constraints. On Libri3Mix and a Dynamic-Mix (2+3) set (speaker count unknown at test time), our model achieves 14.52\%/9.24\% cpWER and 1.95\%/1.12\% WDER, respectively, with GRPO yielding a 35\%/54\% relative cpWER reduction over SFT+CoT.
\end{abstract}

\section{Introduction}

The recent advancements in Speech Large Language Models (SpeechLLMs), such as Qwen-Audio \cite{chu2023qwen}, SALMONN \cite{tang2023salmonn}, and Kimi-Audio \cite{ding2025kimi}, have demonstrated unprecedented zero-shot capabilities in single-speaker automatic speech recognition (ASR) and general audio understanding. However, 
the ``cocktail party problem'' \cite{cherry1953some}, characterized by overlapping speech from multiple talkers, remains a fundamental challenge. Crucially, resolving this acoustic interference requires tackling the joint task of multi-talker ASR and speaker diarization—simultaneously deciphering \textit{what} was said, \textit{who} said it, and \textit{when} it occurred.

To explicitly model this joint task, traditional multi-talker systems often rely on complex, cascaded pipelines involving speech separation, single-talker ASR, and clustering-based speaker diarization \cite{watanabe2020chime}, which are highly susceptible to error propagation. To alleviate this, recent end-to-end models utilizing Permutation Invariant Training (PIT) \cite{yu2017permutation} or Serialized Output Training (SOT) \cite{kanda2020serialized} have significantly simplified the modeling pipeline. Specifically, highly optimized architectures such as the SOT-Conformer \cite{kanda2021end}, Whisper-Sidecar \cite{meng2024empowering} and UME \cite{shakeel2025unifying} have pushed the boundaries of these paradigms, achieving remarkable transcription and diarization accuracy on overlapping speech. While these specialized models excel within their designated frameworks, extending the powerful, reasoning and generative capabilities of foundational SpeechLLMs to this complex scenario remains largely unexplored.


\begin{figure*}[t] 
  \centering
  \includegraphics[width=0.79\textwidth]{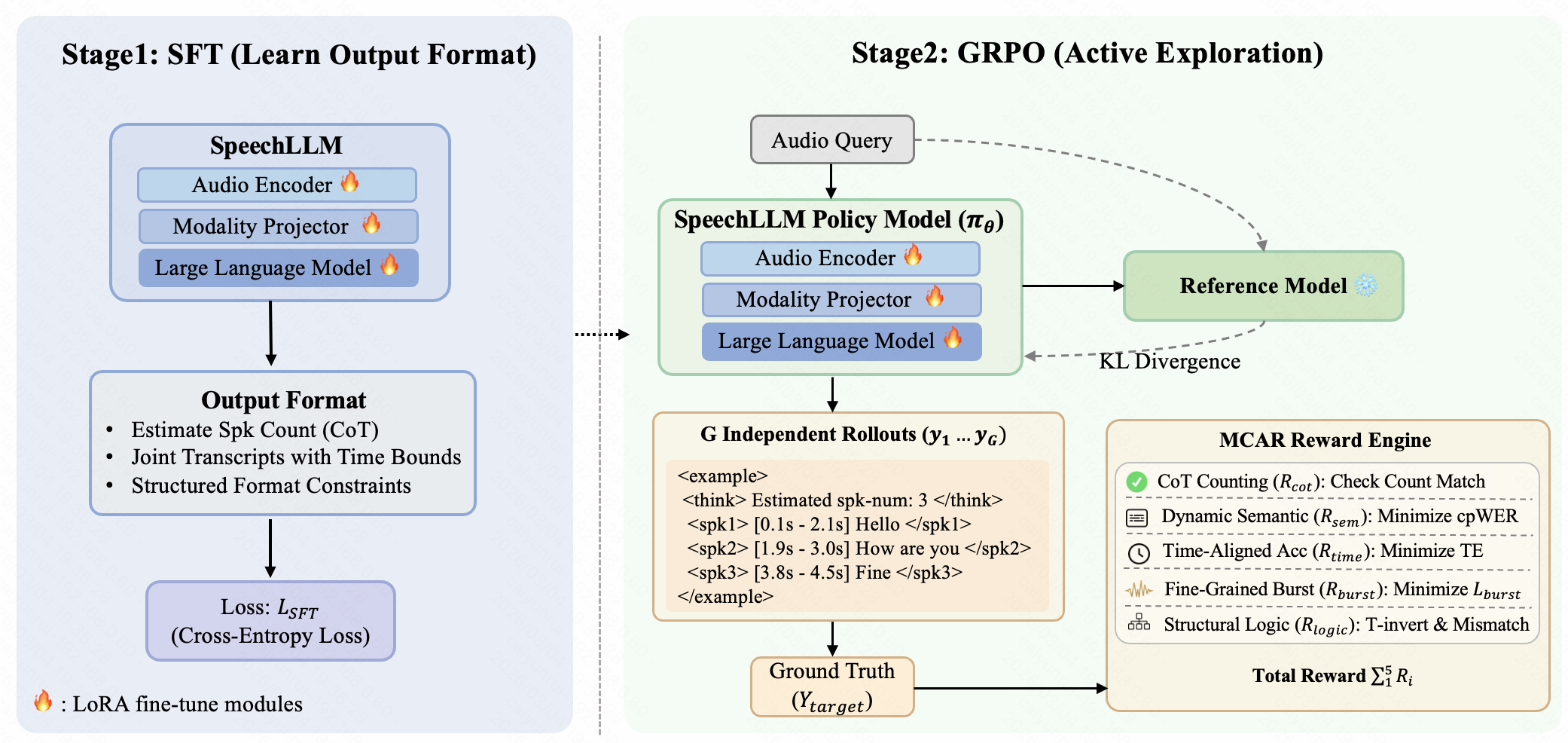}
  \caption{The overall architecture of our proposed two-stage generative framework. \textbf{(a) Stage 1: Constraint-Aware SFT} establishes the output format. \textbf{(b) Stage 2: GRPO with MCAR} shifts the paradigm to active exploration by evaluating multiple independent rollouts against a multi-dimensional reward engine to enforce semantic, temporal, and logical constraints.}
  \label{fig:method_arch}
\end{figure*}


Fundamentally, off-the-shelf SpeechLLMs fail on overlapping speech due to severe acoustic interference. Recent pioneering efforts (e.g., MT-LLM \cite{mengLargeLanguageModel2025}, CMT-LLM \cite{CMTLLMContextualMultiTalker2025}, JEDIS-LLM \cite{shi2025train}) mitigate this by splicing specialized multi-talker audio front-ends into the LLM, typically aligning the modalities via parameter-efficient techniques like Low-Rank Adaptation (LoRA)~\cite{hu2021lora}. To preserve end-to-end elegance, a natural alternative is direct Supervised Fine-Tuning (SFT) to sequentialize overlapping signals via text-driven supervision. However, SFT exhibits critical flaws in dynamic scenarios with unknown speaker counts, and suffers from severe text hallucination~\cite{atwany2025lost} and illogical temporal boundaries~\cite{tan2025enhancing}. These failures are difficult to fix by further SFT: teacher forcing never trains the model on its own malformed generations, while the key objectives are sequence-level and non-differentiable. We therefore adopt reinforcement learning to perform constrained, on-policy exploration over complete hypotheses and directly optimize these task-level constraints.


To break this bottleneck, we propose a parameter- and data-efficient two-stage framework that aligns SpeechLLMs with the physical realities of overlapping speech\footnote{The code is available at https://github.com/caiyunrui/MT-GRPO.}. First we construct a carefully curated, fine-grained SFT dataset. Crucially, we introduce a Chain-of-Thought (CoT) reasoning prefix into the training target, prompting the model to explicitly estimate the speaker count before transcribing. 
Stage 2 introduces Group Relative Policy Optimization (GRPO) \cite{shao2024deepseekmath} to the multi-talker domain, which has recently been explored in simple ASR \cite{shivakumar2025group}. Unlike data-hungry SFT, this phase relies on self-generated rollouts and active exploration. We guide this exploration with a multi-dimensional reward mechanism that directly optimizes Permutation-Invariant WER, rewards CoT-accuracy, enforces structural logic, and penalizes hallucinations. This GRPO alignment effectively shifts the learning paradigm from passive mimicry to constrained exploration, enabling acoustic scene deduction with limited overhead.

Our main contributions are summarized as follows:
\begin{itemize}
\item We are the first to successfully adapt and optimize SpeechLLMs for complex multi-talker scenarios—including 2-speaker, 3-speaker, and challenging dynamic mixtures—using a combined SFT and GRPO training paradigm.
\item We unify Multi-talker ASR and Speaker Diarization into a single end-to-end SpeechLLM reasoning process. Empowered by a fine-grained reward engine, we effectively mitigate the severe hallucination and timestamp degradation issues prevalent in pure SFT models.
\item Experimental results demonstrate that our approach significantly expands the dense-overlap acoustic comprehension capabilities of SpeechLLMs.
\end{itemize}

\section{Methodology}
\label{sec:method}

To tackle the complexities of dynamic multi-talker scenarios, we propose a two-stage adaptation and exploration framework. We first formulate the joint ASR and diarization task via a CoT reasoning mechanism, and subsequently adapt the SpeechLLM through SFT, followed by GRPO guided by a Multi-dimensional Constraint-Aware Reward (MCAR) engine.

\subsection{Problem Formulation}
Given a continuous overlapping speech signal, we extract the acoustic feature sequence $\mathbf{X} \in \mathbb{R}^{T \times F}$, where $T$ and $F$ denote the frame length and feature dimension. The objective of our joint task\footnote{In the context of this work, speaker diarization refers strictly to speaker-attributed ASR and segment boundary detection within highly overlapped short utterances.} is to generate a highly structured token sequence $\mathbf{Y}$, which explicitly captures the dynamic speaker count, identities, temporal boundaries, and spoken transcripts. To prevent severe speaker confusion in dynamic mixtures, we introduce a CoT reasoning prefix $\mathbf{C}$ that forces the model to estimate the active speaker count $N$ before transcription. $\mathbf{Y}$ is defined as:
\begin{equation}
 \mathbf{Y} = \mathbf{C} \oplus \bigoplus_{i=1}^{N} \left[ \langle \text{spk}_i \rangle, t_{s}^{(i)}, t_{e}^{(i)}, \mathbf{W}^{(i)}, \langle /\text{spk}_i \rangle \right] 
\end{equation}
where $\oplus$ denotes sequence concatenation, $N$ is the dynamic number of speakers, $t_{s}^{(i)}$ and $t_{e}^{(i)}$ represent the start and end timestamps, and $\mathbf{W}^{(i)}$ denotes the transcription for the $i$-th speaker. $N$ is unknown and varies per utterance. 

\subsection{Model Architecture \& LoRA Adaptation}
We build our framework upon a foundational SpeechLLM architecture (e.g., Qwen-Audio/Omni~\cite{chu2023qwen,xu2025qwen3} series), comprising an audio encoder $\mathcal{E}_\theta$, a modality projector $\mathcal{P}_\phi$, and a large language model $\mathcal{M}_\psi$. To break the physical limitations of acoustic entanglement in overlapping speech without incurring prohibitive computational costs, we employ a universal Low-Rank Adaptation (LoRA) strategy. Distinct from conventional text-only alignment, we apply LoRA modules universally across all linear projections within the audio encoder $\mathcal{E}_\theta$, the modality projector $\mathcal{P}_\phi$, and the large language model $\mathcal{M}_\psi$. This comprehensive, parameter-efficient adaptation ensures that fine-grained acoustic disentanglement is seamlessly achieved at the front-end, while strictly preventing catastrophic forgetting of the LLM's inherent linguistic and CoT reasoning capabilities.

\subsection{Stage 1: Constraint-Aware Supervised Fine-Tuning}
In the first stage, the primary goal is to establish the fundamental CoT logic and structural templates. To force the SpeechLLM to rely on acoustic features rather than spurious sequential biases, we employ \textit{Permutation-Invariant Prompting}. For each audio sample with $N$ speakers, the transcription segments are randomly shuffled among the $N!$ possible permutations. The model is optimized using the standard auto-regressive negative log-likelihood loss $\mathcal{L}_{SFT}$. While SFT successfully initializes the CoT format and XML-style tags, it intrinsically suffers from exposure bias. When confronted with variable speakers, pure SFT models exhibit continuous hallucinations, temporal degradation, and logical inconsistencies between the predicted CoT count and the actual generated segments.

\subsection{Stage 2: GRPO via Constrained Exploration}
To alleviate the hallucination avalanche and sequence-bias inherent in teacher-forcing, we transition from passive imitation to active exploration via Group Relative Policy Optimization (GRPO), which eliminates the memory-intensive value model by estimating the baseline directly from a group of generated outcomes.

For each audio query $q$, the active policy $\pi_\theta$ samples a group of $G$ independent rollouts $\{y_1, \dots, y_G\}$. The GRPO objective maximizes the expected relative reward using a clipped surrogate function, heavily regularized by the Kullback-Leibler (KL) divergence to the frozen SFT reference model $\pi_{ref}$:
\begin{equation}
c_i(\theta) = \text{clip}(r_i(\theta), 1-\epsilon, 1+\epsilon)
\end{equation}
\begin{equation}
    \mathcal{J}(\theta) = \mathbb{E}_{q, y_i} \Bigg[ \sum_{i=1}^{G} \min \Big( r_i(\theta) \hat{A}_i, c_i(\theta) \hat{A}_i\Big)- \beta \mathbb{D}_\textit{KL}(\pi_\theta || \pi_{ref}) \Bigg]
\end{equation}

where $r_i(\theta) = \frac{\pi_\theta(y_i|q)}{\pi_{\theta_{old}}(y_i|q)}$ is the policy probability ratio, and $\epsilon$ is the clipping bound. The group-relative advantage $\hat{A}_i = \frac{R_i - \text{mean}(R)}{\text{std}(R)}$ is derived from our customized MCAR reward $R_i$, scaling the objective to favor above-average rollouts dynamically. Simultaneously, the KL penalty $\beta \mathbb{D}_{KL}$ anchors the policy, preventing catastrophic deviation from the linguistic and structural priors established during SFT. 

\subsection{Multi-dimensional Constraint-Aware Reward}
The core innovation of our RL stage lies in the MCAR engine. It evaluates the dynamic multi-talker outputs through an exhaustive $N!$ permutation search and enforces five orthogonal physical and logical constraints:

\textbf{1. CoT Counting Reward ($R_{cot}$):} To stabilize exploration early in training, we provide a dense reward if the model correctly deduces the acoustic complexity. We parse the integer $N_{pred}$ from the CoT prefix $\mathbf{C}$ and compare it against the ground truth $N_{true}$. A substantial positive reward ($+1.0$) is granted for an exact match.

\textbf{2. Dynamic Semantic Fidelity ($R_{sem}$):} Standard WER is highly ambiguous when speaker identities are permuted. We implement a dynamic Permutation-Invariant WER (PI-WER, equivalent to the evaluation metric cpWER) that traverses all $N!$ combinations between the predicted and reference segments to find the optimal alignment path. To prevent gradient instability from unbounded errors, PI-WER is mapped into an exponentially decaying reward surface:
\begin{equation}
R_{sem} = \exp(-\gamma \cdot \text{PI-WER}_{N!}(y_i, Y_{ref}))
\end{equation}

\textbf{3. Time-Aligned Accuracy ($R_{time}$):} Based on the optimal alignment path discovered by PI-WER, we compute the absolute Time Error (TE) between the generated and reference timestamps. We tolerate minor deviations ($\tau = 0.2s$) and reward temporal precision via:
\begin{equation}
R_{time} = \exp(-\max(0, \text{TE} - \tau))
\end{equation}

\textbf{4. Fine-Grained Burst Penalty ($R_{burst}$):} 
Acoustic masking often triggers auto-regressive collapse in SpeechLLMs, causing cascading text hallucinations \cite{baranski2025investigation, wang2025calm}. Because standard WER treats scattered and consecutive errors equally, it fails to penalize these localized avalanches. To suppress this, we track $L_{burst}$, the maximum contiguous length of insertion and substitution errors derived from the PI-WER alignment. We impose a severe penalty for ungrounded generation exceeding a tolerance threshold $\tau_b$:
\begin{equation}
R_{burst} = - \lambda_b \cdot \max(0, L_{burst} - \tau_b)
\end{equation}
where $\tau_b$ (e.g., $\tau_b = 2$) permits minor, acoustically plausible misrecognitions, but aggressively penalizes rambling. This explicitly forces the policy to concisely close the speaker tag ($\langle /\text{spk} \rangle$) when acoustic grounding is lost.


\textbf{5. Structural \& Temporal Logic ($R_{logic}$):} We enforce strict adherence to the XML formatting and physical constraints. $R_{logic}$ heavily penalizes missing tags, conversational verbosity outside tags, and logically impossible temporal inversion ($t_s \ge t_e$). Furthermore, severe logical divergence—such as predicting $N_{pred}=3$ in the CoT but generating only 2 speaker tags—triggers a catastrophic penalty.

The final scalar reward $R_i$ is the weighted sum of these five components. This multi-dimensional formulation strictly guides the SpeechLLM to align its generative distribution with the complex, dynamic constraints of multi-talker overlapping speech.

\begin{table*}[t]
\centering
\caption{Evolution of our unified SpeechLLM from Zero-Shot to SFT + CoT + GRPO. The evaluation covers standard 2- and 3-speaker mixtures, as well as the challenging Dynamic-Mix (2+3) setting. Evaluated by cpWER (\%), WDER (\%) ,TE (s) and CoT-Acc (\%).}
\label{tab:internal_progression}
\resizebox{\textwidth}{!}{%
\begin{tabular}{l|ccc|ccc|cccc}
\toprule
\multirow{2}{*}{\textbf{Model Stage}} & \multicolumn{3}{c|}{\textbf{2-Speaker (Libri2Mix)}} & \multicolumn{3}{c|}{\textbf{3-Speaker (Libri3Mix)}} & \multicolumn{4}{c}{\textbf{Dynamic-Mix (2+3 Speakers)}} \\
\cmidrule(lr){2-4} \cmidrule(lr){5-7} \cmidrule(lr){8-11}
 & \textbf{cpWER } & \textbf{WDER } & \textbf{TE } & \textbf{cpWER } & \textbf{WDER} & \textbf{TE} & \textbf{cpWER } & \textbf{WDER } & \textbf{TE } & \textbf{CoT-Acc } \\
\midrule
Qwen2.5-Omni-7B & 58.25 & 8.84 & 2.14 & 81.82 & 25.62 & N/A$^*$ & 69.3 & 18.52 & N/A$^*$ & 32.25 \\
\quad + SFT  & 6.12 & 0.85 & 0.07 & 23.92 & 4.21 & 0.18 & 24.78 & 4.81 & 0.20 & 76.80 \\
\quad + SFT + CoT & 5.83 & 0.67 & 0.05 & 22.45 & 4.03 & 0.18 & 20.17 & 3.73 & 0.15 & 90.06 \\
\midrule
\textbf{+ SFT + CoT + GRPO} & \textbf{4.45} & \textbf{0.48} & \textbf{0.04} & \textbf{14.52} & \textbf{1.95} & \textbf{0.10} & \textbf{9.24} & \textbf{1.12} & \textbf{0.08} & \textbf{99.72} \\
\bottomrule
\end{tabular}%
}
\vspace{4pt} 
\parbox{\textwidth}{\footnotesize $^*$ TE for the zero-shot model is denoted as N/A because the model frequently hallucinates non-numeric characters or unformatted textual boundaries, rendering exact temporal alignment computationally infeasible.}
\end{table*}

\section{Experiments}
\label{sec:experiments}

\subsection{Experimental Setup}

\subsubsection{Datasets} We comprehensively evaluate our framework on the widely adopted \textbf{Libri2Mix} and \textbf{Libri3Mix} datasets \cite{cosentino2020librimix}. Formally, these datasets combine clean utterances from LibriSpeech \cite{panayotov2015librispeech} with WHAM! noise \cite{wichern2019wham} to simulate realistic scenarios. 

To explicitly assess the model's reasoning capacity on dynamic, unknown speaker counts, we further construct a novel \textbf{Dynamic-Mix (2+3)} evaluation set by uniformly sampling and shuffling test utterances from both Libri2Mix and Libri3Mix. We use these benchmarks as controlled diagnostic settings to isolate dense overlap and unknown speaker-count reasoning. Finally, the GRPO policy is optimized on a heavily curated subset of 10,000 highly overlapped speech samples, which are strictly sampled from the official training splits.

\subsubsection{Implementation Details} Our experiments are conducted on the Qwen2.5-Omni-7B foundational model using the MS-Swift framework. We apply Low-Rank Adaptation (LoRA) \cite{hu2021lora} universally across all linear layers within the audio encoder, modality projector, and the LLM, setting rank $r=16$, $\alpha=32$, and dropout $p=0.05$. 

During the SFT stage, the model is trained for 5 epochs with a global batch size of 128 and a learning rate of $2 \times 10^{-5}$ . In the GRPO stage, the learning rate is decayed to $1 \times 10^{-6}$ and $\beta$ is set to $0.04$. For each audio query, the active policy samples $G=8$ independent rollouts, with the clipping parameter $\epsilon$ set to $0.2$. The decay factor in Eq. (4) is set to $\gamma=1.0$. For the fine-grained burst penalty ($R_{burst}$), we set the tolerance window $\tau_b = 2$ tokens and the penalty scale $\lambda_b = 1.0$. The five rewards are equally weighted to compute the total advantage.


\subsubsection{Evaluation Metrics}
To evaluate both transcription fidelity and diarization accuracy, we employ the following metrics:
\begin{itemize}
    \item \textbf{cpWER (Concatenated minimum-Permutation WER):} The primary joint metric for multi-talker ASR, computing the minimum Word Error Rate across all $N!$ speaker-to-transcript permutations. Note that this cpWER is mathematically equivalent to PI-WER utilized in our $R_{sem}$ reward.
    \item \textbf{WDER (Word Diarization Error Rate):} Traditional pipelines rely on frame-level Diarization Error Rate (DER). However, generative SpeechLLMs operate on discrete semantic tokens rather than continuous acoustic frames, making frame-level metrics highly susceptible to boundary quantization artifacts. WDER can evaluate text-level speaker confusion by calculating the percentage of correctly transcribed words assigned to incorrect speakers.
    \item \textbf{TE (Timestamp Error):} The average absolute deviation between the predicted and ground-truth temporal boundaries.
    \item \textbf{CoT-Acc (Chain-of-Thought Accuracy):} Specifically for the Dynamic-Mix setting, measuring the exact-match accuracy of the predicted speaker count prior to text generation.
\end{itemize}

\subsection{Main Results}
\subsubsection{The Evolution of Generative Capability.} 
Table \ref{tab:internal_progression} presents the progressive evolution of our SpeechLLM from zero-shot to the fully RL-aligned architecture. We analyze the results across transcription (cpWER), diarization (WDER), temporal grounding (TE), and dynamic reasoning (CoT-Acc).

Zero-shot Qwen2.5-Omni-7B performs poorly under dense overlap, achieving only 32.25\% speaker-count accuracy on Dynamic-Mix. While SFT injects the multi-talker paradigm, it struggles with complex mixtures. Introducing a Chain-of-Thought (CoT) prefix forces explicit speaker counting, demonstrating the necessity of structural scene deduction prior to autoregressive generation. 

GRPO further improves all metrics, demonstrating the effectiveness of RL alignment beyond passive teacher-forcing. On Libri3Mix, it reduces cpWER to 14.52\% and WDER to 1.95\%. In Dynamic-Mix, CoT-Acc reaches a near-perfect $99.72\%$. Driven by dense logical rewards, the model robustly resolves the global speaker-counting ambiguity before transcribing, driving cpWER down to $9.24\%$.

\subsubsection{Comparison with State-of-the-Art Methods}
Table \ref{tab:sota_comparison} compares our unified framework against specialized architectures and recent SpeechLLMs. Notably, existing generative LLM baselines natively lack explicit speaker grounding, relying on text concatenation without temporal awareness. In contrast, our RL-aligned model naturally bridges this gap without external VAD modules. In terms of transcription, our model achieves a competitive cpWER of $4.45\%$ on Libri2Mix and $14.52\%$ on Libri3Mix, matching or outperforming heavily engineered joint architectures. This confirms that GRPO-aligned semantic deduction serves as a highly effective alternative to hard-coded acoustic components in dense overlaps.

\begin{table}[h]
\centering
\caption{Comparison with state-of-the-art models on LibriMix benchmarks. SD Err. refers to Speaker Diarization Error, where ($*$) denote frame-level DER instead of text-level WDER. Our approach matches or exceeds specialized architectures without relying on external modules.}
\label{tab:sota_comparison}
\resizebox{\columnwidth}{!}{%
\begin{tabular}{l|cc|cc}
\toprule
\multirow{2}{*}{\textbf{Method}} & \multicolumn{2}{c|}{\textbf{Libri2Mix}} & \multicolumn{2}{c}{\textbf{Libri3Mix}} \\
\cmidrule(lr){2-3} \cmidrule(lr){4-5}
 & \textbf{cpWER } & \textbf{SD Err. } & \textbf{cpWER } & \textbf{SD Err. } \\
\midrule
\textit{Specialized Models } & & & & \\
~~Whisper-Sidecar \cite{meng2024empowering} & 4.66 & - & 16.79 & - \\
~~UME \cite{shakeel2025unifying}$*$ & 6.4 & 1.37 & 15.9 & 2.29 \\
~~GEncSep \cite{shi2024serialized} & 6.6 & - & 13.1 & - \\
~~TS-ASR-AD \cite{maeda2025joint}$*$ & 6.61 & 1.23 & 14.81 & 2.66 \\
\midrule
\textit{SpeechLLM (Hybrid)} & & & & \\
~~SOT-LLM \cite{shi2024advancing} & 9.0 & - & - & - \\
~~CMT-LLM \cite{CMTLLMContextualMultiTalker2025} & 7.3 & - & - & - \\
~~SOP-LLM \cite{shi2025serialized} & 3.6 & - & 16.5 & - \\
\midrule
~~\textbf{Ours (Unified)} & 4.45 & 0.48 & 14.52 & 1.95 \\
\bottomrule

\end{tabular}
}
\end{table}

\subsection{Ablation Study on MCAR Constraints}

Table \ref{tab:ablation} isolates the contribution of each component within our MCAR reward engine on the Dynamic-Mix set. The removal of the CoT Counting Reward triggers the most catastrophic degradation. Coupled with the drop in CoT-Acc, this validates that explicitly rewarding global acoustic scene deduction is a prerequisite for accurate autoregressive transcription. Similarly, omitting the Dynamic $N!$ Evaluation disables permutation-invariant optimization, drastically corrupting speaker tracking and causing the cpWER to surge to $16.56\%$. 

Furthermore, disabling the Fine-Grained Burst Penalty degrades cpWER to $11.84\%$, confirming its critical role in suppressing localized hallucination avalanches and auto-regressive collapse. Finally, removing the Time-Aligned Accuracy or Structural Logic rewards yields minor but consistent regressions. This indicates that while they are not the primary drivers of semantic accuracy, these components act as essential regularizers, rigidly enforcing temporal formatting and preventing structural degradation during active RL exploration.

\begin{table}[h]
\centering
\caption{Ablation study of MCAR components on the Dynamic-Mix set. Metrics highlight the degradation when specific rewards are removed.}
\label{tab:ablation}
\footnotesize
\setlength{\tabcolsep}{4.5pt}
\begin{tabular}{l|ccc}
\toprule
\textbf{Reward Configuration} & \textbf{cpWER } & \textbf{WDER } & \textbf{CoT-Acc } \\
\midrule
Full CoT + GRPO (Ours) & \textbf{9.24} & \textbf{1.12} & \textbf{99.72} \\
\midrule
\textit{w/o} CoT Counting Reward &18.02 & 3.13 & 91.67 \\
\textit{w/o} Fine-Grained Burst Penalty  & 11.84 & 1.32 & 98.50 \\
\textit{w/o} Time-Aligned Acc  & 9.66 & 1.20 & 99.20 \\
\textit{w/o} Structural Logic  & 9.85 & 1.17 & 99.67 \\
\textit{w/o} Dynamic $N!$ Eval  & 16.56 & 2.84 & 99.17 \\
\bottomrule
\end{tabular}
\end{table}

\section{Conclusion}
In this paper, we introduced a two-stage generative framework that empowers SpeechLLMs to multi-talker overlapping speech tasks. By integrating Chain-of-Thought (CoT) reasoning with Group Relative Policy Optimization (GRPO), our model transitions from passive text mimicry to active acoustic scene deduction. Evaluated on the LibriMix and Dynamic-Mix benchmarks, this unified approach achieves competitive cpWER and diarization precision (WDER) without relying on external VAD or clustering modules. 
For future work, we plan to scale this RL paradigm to larger real-world conversational datasets like AMI and AliMeeting. We will also explore finer-grained reward mechanisms to enhance temporal resolution and extend this generative approach to broader tasks.
\newpage
\section{Acknowledgements}
This study was supported by the Centre for Perceptual and Interactive Intelligence, a CUHK-led InnoCentre under the InnoHK initiative of the Innovation and Technology Commission of the Hong Kong Special Administrative Region Government.
\section{Use of Generative AI Disclosure}
During the preparation of this manuscript, the authors used generative AI tools exclusively for the purpose of language editing and manuscript polishing to improve readability. These tools were not used to generate any core scientific ideas, experimental data, or technical contributions. All authors have thoroughly reviewed and approved the final version of the manuscript, and assume full responsibility for the integrity and entirety of its content.
\bibliographystyle{IEEEtran}
\bibliography{mybib}

\end{document}